Proceedings of the 2021 Pre-ICIS SIGDSA Symposium

Special Interest Group on Decision Support and Analytics (SIGDSA)

12-2021

# The Competitive Leverage Paradox Effect on Information Systems Life Cycle

Samaa Elnagar
*Virginia commonwealth university*, elnagarsa@vcu.edu

Kweku-Muata Osei-Bryson
*Virginia Commonwealth University*, KMOsei@vcu.edu

# The Competitive Leverage Paradox Effect on Information Systems Life Cycle

**Samaa Elnagar**
Virginia Commonwealth University,
Richmond, VA, USA
elnagarsa@vcu.edu

**Kweku-Muata Osei-Bryson**
Virginia Commonwealth University,
Richmond, VA, USA
kmosei@vcu.edu

*Completed Research Paper*

## Abstract

The fierce market competition has put pressure on organizations leveraging their value chains. The continuous development in strategic technologies such as Artificial Intelligence (AI) has pushed organizations to continuously acquire new Intelligent Information Systems (IIS) while underutilizing existing ones leading to the competitive leverage paradox. However, research on underutilizing IIS has focused on the social and organizational aspects of the problem, ignoring the flaws in designing and evaluating IIS. One of the overlooked factors is the effective life span of an IIS. This research conducted a systematic literature review to profoundly investigate the determinants of the competitive leverage paradox and its effect on the IIS life cycle. The research studies the IISs from economic and design perspectives. We also explore the design and strategic factors that led to defects in the effective life cycle of IIS. This research calls to consider the economic, and design factors in addressing the underutilization of IIS. The study also presents future research propositions to enhance IIS life cycle and return on investment.

### Keywords

Competitive Leverage, Value Chain, Underutilization, Economics of Information Systems, Information System Life Cycle, Return on Investment, Effective Lifespan, Artificial Intelligence.

## Introduction

The trend toward data-driven decision-making has created an exciting paradigm shift in how organizations create and leverage knowledge for decision-making (Davenport and Kudyba 2016). According to Gartner's 2018 technology trend survey (Li et al. 2018), Artificial Intelligence (AI) is the No. 1 strategic technology. An Intelligent Information System (IIS) can be considered to be an Information System (IS) that is based on AI (Aronson et al. 2005; Elhoseny et al. 2017; Ghoshal and Moran 1996). The organizational value chain no longer begins and ends at the organization itself, but customers, organizations, and global markets are indivisible parts of the chain (Zhaohao et al. 2012). We can depict the organizational value chain as a braided value chain in which several value chains are interconnected, and each chain adds more strength to the whole chain. The organizational value chain encompasses many specialized value chains such as knowledge value chain (Chyi Lee and Yang 2000), innovation value chain (Hansen and Birkinshaw 2007) and supply chain. AI-based systems enable the addition of new values to the organization's value chain and generate disruptive innovations that can significantly change existing value propositions (Huang et al., 2017). However, a study by Davenport et al. (2018) found that while executives believe that AI will substantially transform their companies within three years, more than two-thirds of the IISs are underutilized (Davenport and Ronanki 2018). Reports included little perceived quality and high costs of implementing many IIS (Elnagar and Osei-Bryson 2020).

Despite the current AI revolution, technology outsourcing, and governance practices, most IISs don't pay off the cost and risk associated with them. Due to the increase in economic and competitive pressure, many organizations continuously underutilize existing IISs while constantly attaining or outsourcing new ones to gain more competitive leverage. However, attaining new IISs result in a continuous increase in cost and decrease in *Return on Investment (ROI),* developing a *competitive leverage paradox.* We argue that the *competitive leverage paradox* is a key contributor to IIS underutilization and the continued economic loss.

While IS infusion research addressed the utilization issue extensively (Tams et al. 2018), we believe that IS research is focusing only on identifying the social and organizational problems associated with IIS underutilization that will not provide a viable solution to the problem. Most IISs infusion research discards the economic and design factors in solving the underutilization problem (J. Hester 2014). After identifying the previous literature on the underutilization of IIS, we found that many economic and design aspects are strongly correlated with IISs underutilization.

To get a broader depiction of how the economic and design issues contribute primarily to the underutilization of IIS, we must scrutinize the design of an archetypical IIS while paying attention to the organizational inertia and examining value path dependencies. We also have to trace the IIS life cycle, identify the grounds of underutilization, and quantify what is expected versus the actual outcome. In addition, we critically discuss the opportunities for behavioral, design science, and economics of IS research to build an effective IIS that continuously adds value to the organizational value chain. We believe that the issues raised by this research are highly disruptive to the design research process, which requires re-assessment of many IS status quo research practices, methodologies, and theories.

## The Competitive Leverage Paradox

Organizations have made significant investments in implementing different information systems (IS). In organizational settings, a disruptive phenomenon could significantly alter value chains (Thomond and Lettice 2002). However, more than half of these IS implementations have reported failures due to underutilization(Ng and Kim 2009). Davenport et al. conducted a study of more than 152 projects. They found that while executives believe that AI will substantially transform their companies, more than two-thirds of the projects failed (Davenport and Ronanki 2018). Justifying why some IS artifacts succussed, whereas others fail, is a multifaceted and multi-factorial problem. The underutilization of IIS incorporates many theoretical perspectives, such as Practice theory (Cecez-Kecmanovic et al. 2014), transaction cost theory (Williamson 1993), Socio-technical systems theory (J. Hester 2014), and organization transformation theory(Quinn and Cameron 1988). The factors affecting the use, impact, success, and failure of information systems have been studied extensively (Hughes et al. 2019). DeLone defined IS success actors as System Quality, Information Quality, Use, User Satisfaction, Individual Impact, and Organizational Impact (DeLone and McLean 1992).

IS failure can be defined as "either the implemented system not meeting the user expectations or inability of creating working or a functioning system." Despite the efforts to understand the underlying success factors, IS failure rate remains stubbornly high (Dwivedi et al. 2015). Deng et al. revealed seven emergent constructs underutilizing system use, including long training time, data, workflow, role authorization, users' lack of knowledge, system error, and user-system interaction (Deng and Chi 2012). Although IS infusion is required for realizing expected Returns On Investments (ROI), most IS research has focused on the initial adoption and continuance with only a handful examined infusion studies (Karimi et al. 2007). These few IS infusion studies have produced inconclusive results as they have employed models and factors used for adoption and continued use (Bacon 1992).

One of the results of IIS underutilization is economic loss. However, most research discards the economic factor in addressing the cause of underutilization. There have been many debates whether the financial criteria are adequate to evaluate IS investments, given the intangible and strategic nature of some of the IS benefits (Badiru 1990). Switching costs were found to be a primary factor for underutilization. Switching costs is the disutility associated with switching the increase in inputs (e.g., more time and effort to do the work) and the decrease in outcomes (Chen and Hitt 2002).

Based on status quo bias theory, switching costs consist of transition costs, uncertainty costs, and sunk costs. Sunk costs refer to previous commitments, which cause reluctance to *switch to a new alternative.* Transition costs are the costs incurred in adapting to the new situation (Kim and Kankanhalli 2009). switching costs increase user resistance both directly and indirectly through their effect on perceived value (Samuelson and Zeckhauser 1988). Therefore, our main hypothesis that *the economic and strategic loss* associated with the transition cost, the cost of the new IIS, and desertion cost of the old IIS might outweigh the competitive leverage supposed to be gained by the new IIS.

# Literature Review

We undertook a critical structured (systematic and iterative) literature review to explore the determinants for the underutilization of IIS and exploit the findings to assert the competitive leverage paradox phenomena. Based on the literature review findings, we could build a framework that provides in-depth analysis for IIS underutilization. In addition, the literature review findings should reinforce our positions to include the economic factor in the design and evaluation of IIS. We followed the literature review methodology developed by Wolfswinkel (Wolfswinkel et al. 2013) to establish rigor in reviewing the literature.

## *Define the scope*

The scope of the literature is performed according to Wolfswinkel methodology. The literature review includes four basic steps to (1) define the objective(scope) of the review, (2) search, select, and discard irrelevant literature, (3) analyze the selected corpus, and (4) discuss the findings. The research design of the literature review is represented in figure 1.

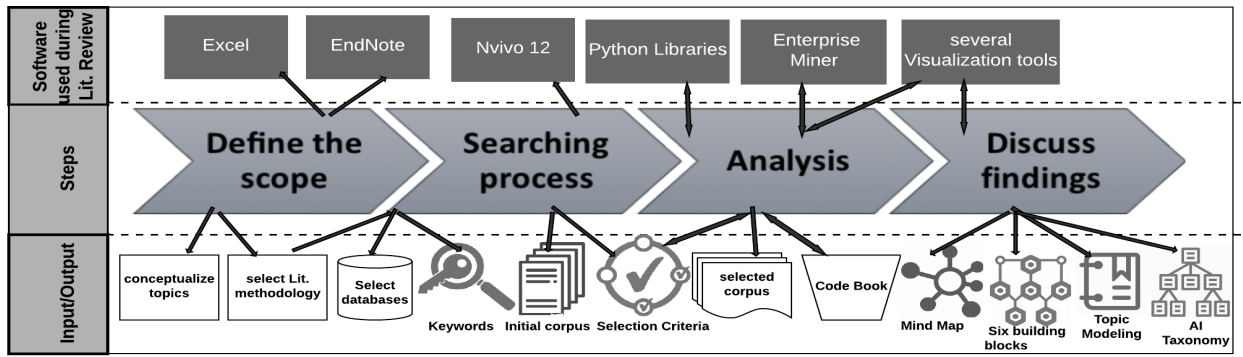

**Figure 1. Research Design of the Literature Review**

## *Searching Process*

Due to the interdisciplinary nature of the problem, we conducted the review on both academic publications, professional publications, practice-based research, and market reports to comprehensively examine the economic, technological, social, and strategic factors of the problem. However, without including market reports and reviews, it would be hard to depict the big picture of the problem using the academic literature only. In addition, previous research on underutilization neither addressed the economic and design factors nor considered the life cycle of the IIS. Therefore, it was necessary to extend the literature beyond IS research to capture the praxis implications in economic reports of underutilization. This knowledge is rarely to be found in academic publications.

### Selection Rejection Criteria

The selection/ rejection criteria were guided by multiple academic sources (Torraco 2005; Vom Brocke et al. 2015) which is (1) the relevance to the topic of IS underutilization, IS failure, and economic evaluation of IS (2) the article should be empirically grounded research such as case and field study. We chose articles where IISs were thoroughly evaluated, and we discarded abstracted frameworks. In addition, we discarded any business report with a pure economic focus, and we chose only technology-related reports. We discarded any review paper that doesn't provide any current limitations.

The search process was conducted in two rounds. In the first round, we reviewed articles and saved the imperative knowledge on an excel file to ease different analysis processes. Within each article, we searched for the technology used, the current gap in the literature, evaluation methodology, limitations of the system, future work, and reflection on IS strategizing. In the second round, we conducted an in-depth systematic literature review (Jones and Gatrell 2014) that compromised backward searching and forward-searching (Webster and Watson 2002). In backward searching, we scrutinized where the author cited relevant references to our research corpora. We used forward-searching with articles that we believe they are highly relevant by searching in the "cited by" of Google Scholar and Web of Science. The "cited by" or forward search for newer publications ensure the currency of our literature review. We created a compilation of references in an EndNote library that is available upon request.

## *Analysis*

Our analysis focused more on identifying the supporting evidence for the underutilization of IIS based on the economic and design issues (Jones and Gatrell, 2014). We aimed to provide new insights that can contribute to future research and, thus, to go beyond merely mapping or describing the current discourse. To this end, we engaged in an iterative process of open and selective coding for synthesizing insights from the selected literature.

**Exploratory Analysis:** To gain a better decomposition of the problem, we formulated the codes resulted from the NVivo coding (Thematic literature annotation) into the major strategic dimensions of the problem.

**Exploitive Analysis:** We will use topic modeling to identify if the literature asserts that the competitive leverage paradox phenomena contribute to the underutilization of IIS. We used topic modeling to comparing generated concepts with the manually developed Nvivo codes. This comparison should confirm the validity of the exploratory Analysis

## *Findings*

### Exploratory Analysis findings

By grouping codes into coherent correlated groups, we ended up having six groups of codings: AI, knowledge, IS, Value Chain, Economics, Organization. The preliminary coding output of the explorative analysis is represented as a mind map, as shown in Figure 2. After identifying the previous literature on the underutilization of IIS, we found that the economic factor is strongly correlated to the underutilization of existing IISs and adopting new alternatives. IIS were considered business assets and a strategic solution that adds economic values to the organizational value chains. Underutilization was also coined with technology currency used in existing and misalignment with the organizational strategic goals.

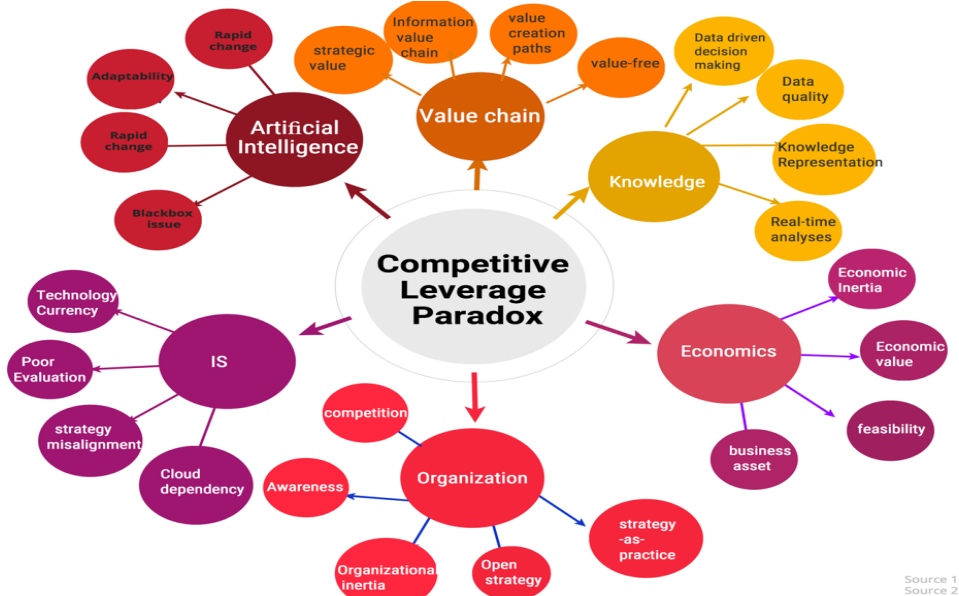

**Figure 2. Preliminary Exploratory Research Finding**

### The Exploitive Analysis

The exploitive analysis was conducted by putting together technology used, current gaps in the literature, evaluation methodology, limitations of the system, future work, and reflections on IS strategizing to be stored in an *Excel* file. This *Excel* file was used as an input corpus for Natural Language Processing (NLP) using Python. We conducted topic modeling analysis using the Python 3 libraries of SpaCy's NLTK, and pyLDAvis. The preliminary output of topic modeling has shown coherence between the modeled topics and the determinants of the competitive leverage paradox. As shown in Figure 3, topic number one, marked in red on the left side, incorporates the organizational strategy and capability. The second topic on the left represents the application risks associated with AI. The fourth topic is related to the investment of AI, and the fifth is associated with the technical aspects of AI. Hence, we can conclude that the main dimensions of the problem are organizational strategy and capability, or the Relationships between Investments in AI & resulting value.

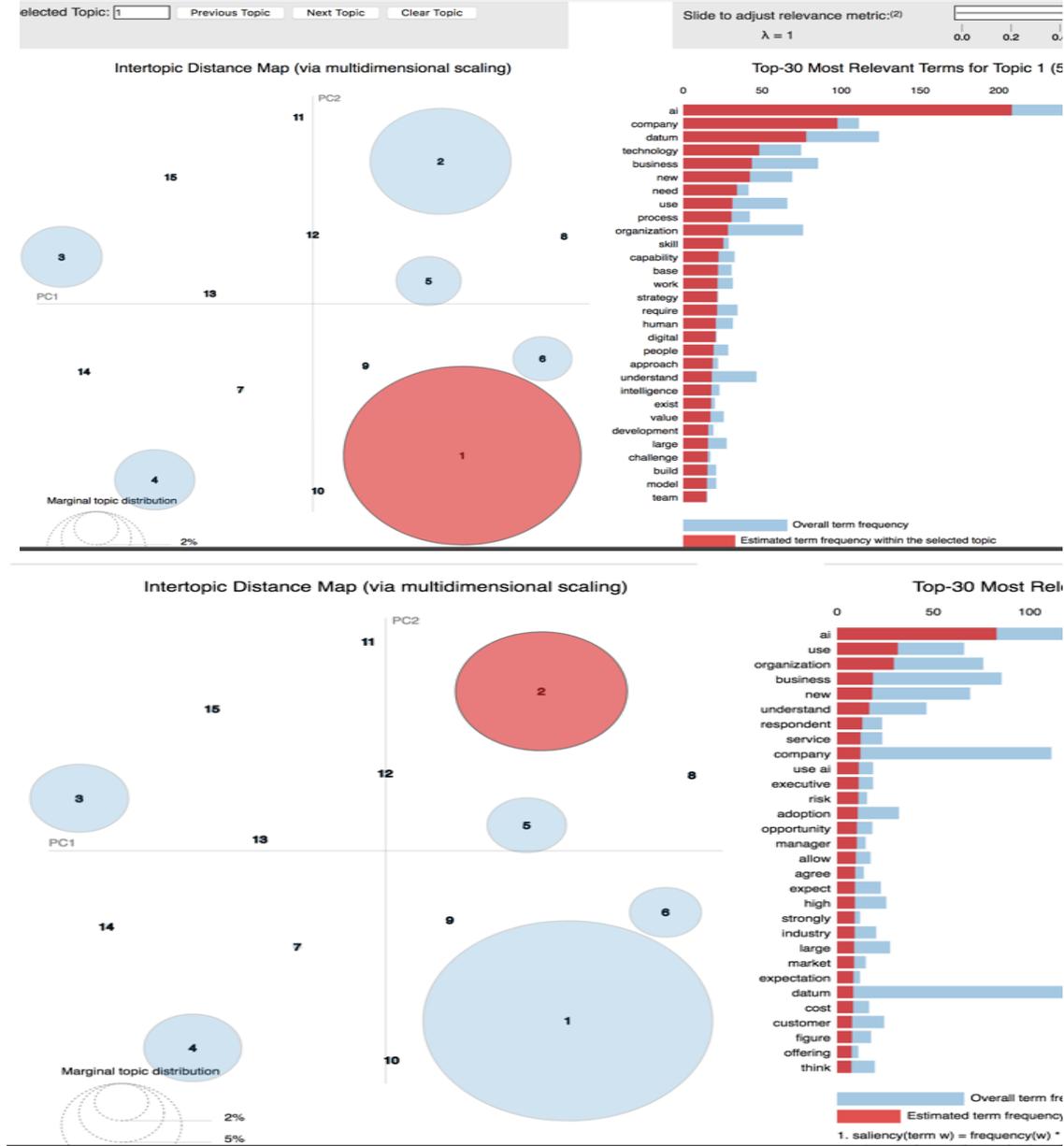

**Figure 3: Topic Modeling results.**

## The Organizational Life Cycle of Information Systems

To get a broader depiction of how the economic factor contributes primarily to the underutilization of IIS, we followed the IIS organizational life cycle in the ideal case of application (where IIS matches the design expected output) (Sebastian et al. 2017). When we study the IIS life cycle, we must consider that conventional AI systems cannot evolve automatically, and they cannot adapt to all changes. So, we should expect an optimal point that after a period of time, IIS cannot develop further. We will call this point (*the maximum expected outcome*). After this point, the output of the system functions remains constant and adds no new value. In the future, systems could continuously evolve and adapt, we should not expect any steady points. On the other hand, the system functions should keep evolving according to changes in the environment (Spohrer and Banavar 2015).

IIS organizational life cycle is represented in figure 4. In the ideal case, when the company first adopts a new IIS, we should not be expecting a linear increase in Return on Investment (ROI). On the contrary, there is a decent amount of time that the ROI will be below the normal *equilibrium point* (ROI before IIS has been introduced) or the *neutral equilibrium,* as Herbert Simon has defined (Simon, 1991). On the left side of figure 4, we will notice a *steady negative* ROI. In fact, the IIS is still not fully operating because of switching cost, sunk cost, user resistance, and organizational inertia. In addition, the cost of operating and depreciating cost outweighs the output generated by IIS. After a while, the IIS will result in increasing the ROI linearly, adding a positive ROI value. In this period, the generated outcome from IIS is greater than the cost associated with it until *the maximum expected outcome* is reached (the expected ROI while designing the IIS).

After *the maximum expected outcome* is reached, the system is expected to have a stable performance and generate steady ROI. We refer to the total time between the beginning of the IIS application until it reaches the *maximum expected outcome* as the *organizational life cycle* of IIS. This life cycle takes the shape of the hyperbolic tangent function $Tanh$, which is known to solve complex infinity problems (Weisstein 2002). We can argue that $Tanh$ has two parts centered towards the *neutral equilibrium*, where the effective reflection of the IIS is in the positive part in the first quarter.

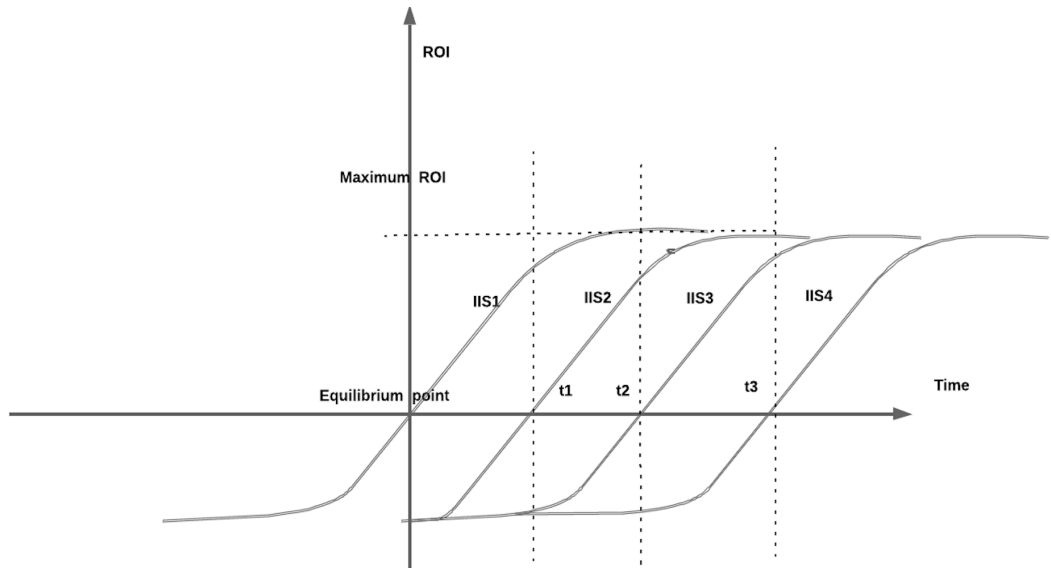

**Figure 4: Organizational Life Cycle of Information Systems**

## The Infinite Loop of Competitive Leverage

*So is this life cycle is the actual life cycle?* The answer is no. usually, organizations underutilize existing IIS before reaching the *maximum expected outcome (Deng and Chi 2012)*. In fact, organizations start adopting new IIS at or before the *equilibrium point,* or before the IIS produces any new value. Fueled by market pressure; outdated technology used in current IIS; and unawareness of IIS capabilities, organizations continue investing in new IISs to increase their competitive leverage.

As we can see in Figure 4, when the organization started investing in IIS2 at the equilibrium point of IIS1, the ROI decreased again because of the cost associated with adopting IIS2. We can notice from the figure that the maximum ROI didn't change because the initial cost associated with investing in IIS2 never exceeded this *maximum ROI* point of IIS1. Repeatedly with investing in IIS3 and IIS4, the maximum ROI never exceeded the maximum ROI of IIS1 if not decreased, in fact. A more interesting finding is that at any point t1, t2, or t4, the sum of ROI is actually **zero** (*the equilibrium point*). Even if IIS1 output increases ROI, the cost associated with IIS2 and IIS3 outweighs the positive increase in IIS1.

So we can represent this competitive leverage paradox as an infinite loop that begins and ends at the same point forcing organizations to invest more money to gain competitive leverage without any reflection on ROI. Figure 5 gives an illustration of the competitive leverage infinite loop. The loop begins with the pressure of the market and rapid change in business needs, such as making complex decisions. The organization suffers from a deficiency in the existing system and outdated technologies, which call for new IIS to solve the new business problems. Then, organizations keep investing with no actual refelction on ROI. While in fact, the organization achieves on the edge performance (*equilibrium point*). For example: an organization planned to use/deprecate an IIS for 10 years. However, after three years, the organization acquired a new IIS to cope with market needs. Therefore, companies should weigh the cost of a new IIS and the actual leverage added value.

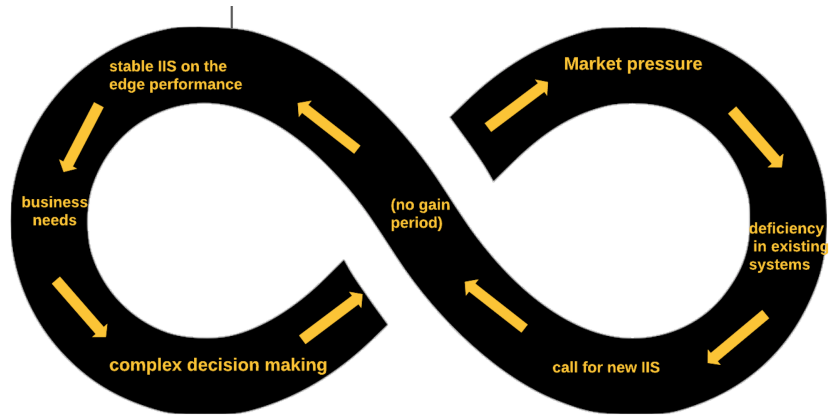

**Figure 5: the competitive leverage infinite loop**

## Reflection on Strategic Information Systems

We reflected on the issues developed from the literature to summarize the determinants for the underutilization of IIS. However, the economic codes were equally related to three dimensions: IS research, market, and organization. Also, we added the economic issues of the strategy provided by (Evans and Wurster 1997). By enclosing the issues resulted from the codes into the three dimensions, we have framed the determinants of the underutilization of IIS, as shown in Figure 6.

Information systems research could strongly contribute to solving the competitive leverage paradox. Designing information systems has suffered from the lack of proper evaluation metrics and overlooking important design guidelines, and eventually, those overlooked metrics will be the challenges faced by IISs (Margaria and Steffen 2006). Moreover, intelligent systems are not evaluated at all or evaluated using subjective qualitative measures (Gretzel 2011). The focus of current IISs evaluation is on the general aspects of information system success, involving measures such as intention to use or actual use and user satisfaction (DeLone and McLean 2003).

Design theories that guide designing IS artifacts have been subjects of debate. Although theories are essential in designing DS artifacts, limiting the design to a confined theoretical perspective, trying to fit the design to a specific theory has deviated the IIS from their anticipated functionality (Merton & Merton, 1968). Otherwise, a theory should not be a rigid formation but a malleable architecture that changes according to design settings and outputs. Studies have found that kernel theories are at such a high level of abstraction that their relationship to design is challenging to discern. In addition, technologies used in building information systems are outdated and don't reflect the current scientific advancements (Elnagar and Thomas 2020).

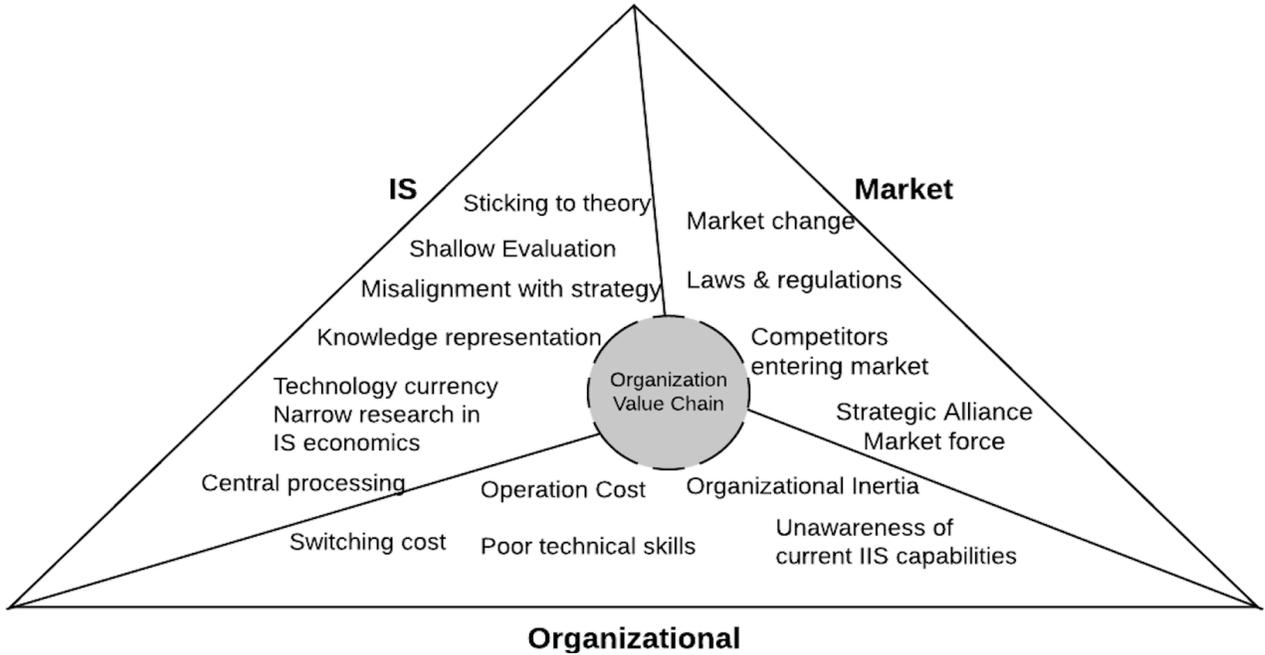

**Figure 6: Determinants for the underutilization of IIS.**

Market pressure is inevitable. The rapid market change in most business sectors created hardship to compete in the market. Competitors market entry, frequent changes in the value chains, and compliance to new regulations and laws have narrowed the chance to gain competitive leverage over rivals (Kim et al. 2017). Therefore, forecasting of upcoming market changes is necessary before investing in a long-term asset such as IIS.

Every organization is a unique structure that an Information System must align with its strategy. Therefore, defining the scope of designing an IIS must gain priority. The design of a local-level artifact created for a specific context should be the dominant approach through a combination of action and design research (Sein et al. 2011). Before investing in IIS, organizations must solve the organizational inertia and consider the cost related with adopting a new system. The employees should have the required technical skills and are able to use the system fully before the system is actually implemented.

## The Building Blocks for Effective IIS

Based on the explorative and exploitive analysis output, we can formalize the main strategies that need to be addressed carefully for designing an effective utilized IIS is discussed below in figure 7.

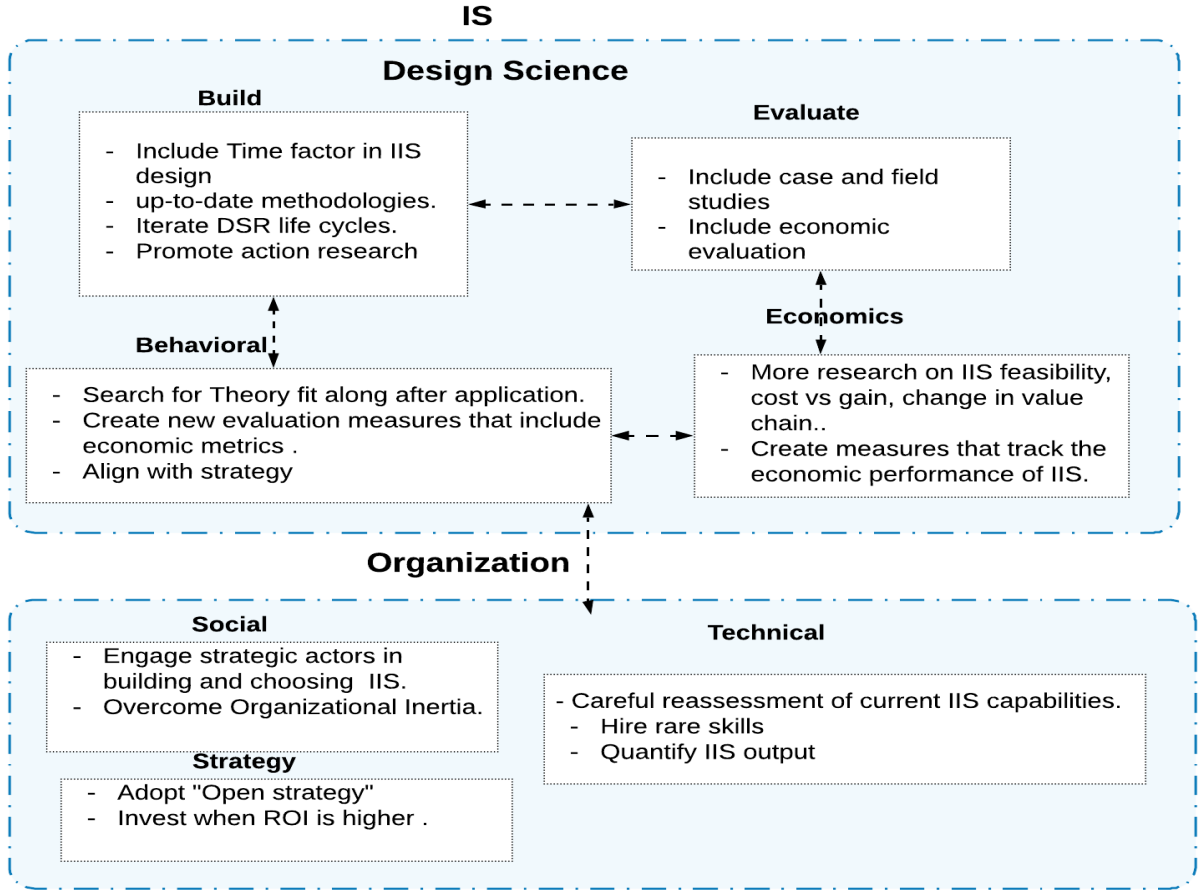

**Figure 7: The Six Building Blocks for effective design of IIS.**

In the guidance of contemporary design science research genres (Rai 2017), we can solve the IS dimension problems of the competitive leverage paradox as: *the economic dimension, the design dimension, and the evaluation dimension.*

The economic dimension of the problem calls for research on the cost vs. gain of information systems (Hester 2014). Organizations should define how investing in an information system will pay back the associated risks and costs. While it is challenging to expect how long an information system will be used effectively, there should be a minimum expected life span associated with detailed instructions on how to decrease the *steady negative* ROI to reach the *maximum expected outcome* faster.

The problems in the design dimension are strongly related to the technologies used to build an IIS, the design methodology, and the scope of the design. At first, the early design of the system should use up-to-date technologies in the field. In the case of designing an abstract artifact that is expected to serve multiple organizations, the scope of the design should be adjusted to fit each organization's strategy. In addition, the role of theory should be as a design input as well as a design output that are accumulated along with the system design (Kuechler and Vaishnavi 2012). Using Agile-based methods is a solution to avoid organizational inertia and unfamiliarity with new systems. Applying the methodology of Strategy-as-practice or "3P" (practitioners, practices, and praxis) (Tavakoli et al. 2017) is one of the promising approaches where strategic users are critical designers of the system.

The Evaluation of Information Systems is the most critical dimension. Here comes the role of IS to create representative evaluation measure that effectively measure the technical and economic impacts of the system and calculate the minimum expected lifespan for each designed IIS. That will help organizations estimate the ROI and decide effectively to invest in an IIS.

From the organizational side, organizations should be aware that technical fit is not the only dimension to be considered while investing in an IIS. Other social and strategic measures should be considered. The strategic actors that will be using the system should be part of the investment decisions. As discussed earlier, unawareness of the problem dimensions and the ineffective use of IIS, leading to underutilization. Adopting an open strategy where all users are engaged in the testing and evaluation of and IIS will help reach the equilibrium point faster, as in figure 4.

## Preliminary Propositions and Future Research

We could recapitulate the issues discussed earlier as preliminary propositions towards solving the competitive leverage paradox and expand the organizational life cycle of information systems as in table 1.

| Proposition | Example of Future Research Considerations |
|---|---|
| Support further studies of the economic gain versus cost in the design of IIS. | How could research in economics of IS enhance IIS utilization? |
| Adopt iterative DSR cycles and adaptive DSR that produce alternative designs. | How creating candidate designs that are based on key valuation criteria will affect the value of created IT systems? |
| Focus on design theories development instead of theory fitting. | How applying new design strategies such as strategy-as-practice decrease the steady negative ROI of IIS life cycle? |
| IIS should be designed to create positive strategic impacts. | How will apply an open strategy, build an IIS that meets the organization's strategic needs? |
| Adopt the most current technologies to avoid outdated technology issues. | How would applying the most current AI technologies enhance the *IIS lifespan*, increase the ROI and increase the maximum expected outcome? |

**Table 1: Preliminary Propositions and Future Research to The Competitive Leverage Paradox.**

## Conclusion

Many IIS notoriously cost draining with little perceived effectiveness and engendered financial and strategic organizational consequences. The preliminary systematic literature review of different IIS revealed that IS research undermines the role of the economic and design factors in addressing IIS underutilization leading to the competitive leverage paradox. Idiosyncratically, there is a crucial call for a paradigm shift in building and evaluating IT artifacts that balance between the economic, organizational and the IS design dimensions. First, we recommend promoting action research and empirical studies as effective methods to increase the alignment with organizational strategy and market goals. Second, we argue that we need fewer theories before design and more theory after design (Law and Urry 2004). Third, the paper calls for paradigmatic and practical solutions that extend the traditional scientific model of research (Gregor and Klein 2014) to effective viable solutions. Fourth, we need to reinforce the design of IIS based on DSR guidelines raised by Arun Rai (Rai 2017).

We establish the *practical relevance* of this research to strategic IS (SIS) by calling researchers to effectively engage with practitioners towards convergent IS-business alignment and effective IS planning (Moeini et al. 2019). In addition, the research synthesize various technical, economic, and organizational strategies that construct effective IISs conducive to adding value to the organizational value chain (Toffel 2016). The scope of this research is not limited to *IS Design Science* but also to *strategic management*. Revisiting the underutilization problem from the lenses of the economic, operational, and design perspectives reveals many organizational and IS strategic issues to be approached. Without a paradigmatic solution for this problem, new-generation AI might not achieve the critical reflection on organizations requirements leading to serious economic and strategic failures. Future propositions offered by the paper could inspire new venues of IS research.